# Gate-Controllable Quadri-Layertronics in 2D Multiferroic Antiferromagnet

Ting Zhang,[1*] Mingsheng Wang,[1] Xilong Xu,[2] Ying Dai,[3] Yandong Ma[3*]

[1]School of Physics and Technology, University of Jinan, Jinan 250022, People's Republic of China

[2]Department of Physics, Washington University in St. Louis, Missouri 63130, USA

[3]School of Physics, Shandong University, Jinan 250100, People's Republic of China

**Abstract**

Layertronics that manifests layer Hall effect is typically considered to intrinsically possess binary physics. Using symmetry arguments and a low-energy $k·p$ model, we show that the layer physics in layertronics can be engineered into quaternary mode, giving rise to the concept of quadri-layertronics. The mechanism correlates to the interplay between out-of-plane ferroelectricity and valley physics in antiferromagnetic multiferroic quadrilayer, which enables the layer-locked Berry curvature and Hall effect, i.e., deflecting the carriers with four different layer physics to move in specific directions. More importantly, the quadri-layertronics can be generated and manipulated by controlling the interlayer dipole arrangements via a gate voltage, allowing for the selective induction and detection of layer Hall effect in specific layers. Using first principles calculations, we further demonstrate the gate control of quadri-layertronics in multiferroic antiferromagnet of quadrilayer $OsCl_2$. These explored phenomena and insights greatly enrich the research on layertronics.

**Keywords**: quadri-layertronics; valley; multiferroic; sliding ferroelectricity; first-principles.

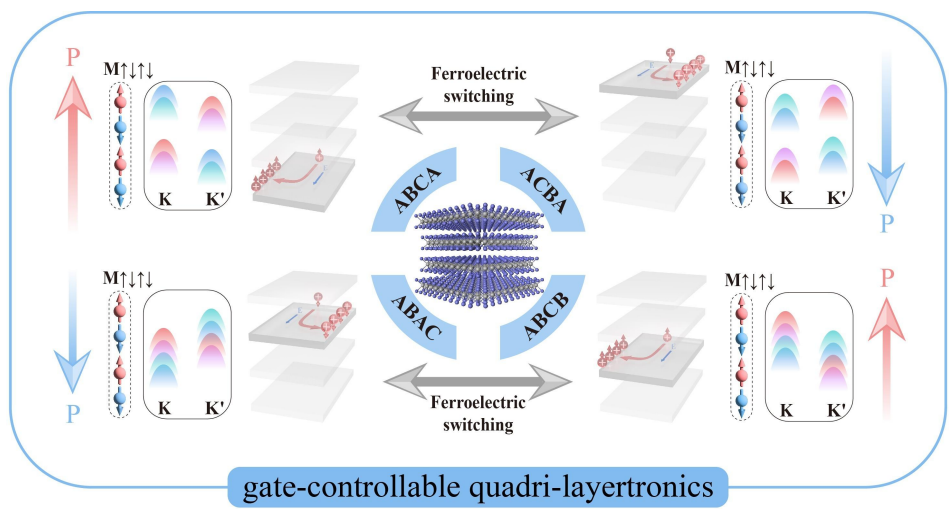

**Introduction**

Berry curvature, describing the local entanglement of Bloch electrons, is a fundamental quantity in condensed-matter physics, particularly when symmetry is broken[1]. In crystalline solids, its interplay with microscopic degrees of freedom can yield intriguing phenomena. Typical examples includes the quantum anomalous Hall effect (QAHE)[2–6] and valley Hall effect (VHE)[7–13]. While the former intertwines Berry curvature to the spin degree of freedom associated with broken time-reversal ($T$) symmetry[14,15], the later couples it to the valley degree of freedom under broken space-inversion ($P$) symmetry[10,16]. Due to the intrinsic symmetry constrain, QAHE and VHE typically harbors binary physics, which is promising for encoding information. At present, with the rise of 2D materials, these Berry curvature driven Hall effects have attracted broad interest in both fundamental research and practical applications[17,18].

In addition to spin and valley degrees, Berry curvature is recently demonstrated to be able to couple with the unique layer degree of freedom[19–21]. This coupling can host the layer Hall effect (LHE) of layer-polarized anomalous Hall effect (LP-AHE), i.e., electrons in different layers are deflected in opposite directions, leading to the concept of layertrinics. Obviously, layertronics is prominent to advancing physics and high-performance information processing and storage, and thus has generated extensive interest[22,23]. The layertronics was first observed in antiferromagnetic (AFM) axion insulator $MnBi_2Te_4$[19]. The past years have seen impressive progress in realizing valleytronics in two-dimensional (2D) AFM and ferromagnetic (FM) systems[24–28]. Despite the significance of layertronics, akin to other Berry curvature related Hall effects, layer Hall effect is also generally considered to be constrained with binary physics. Whether ternary or quaternary physics can be realized in Berry curvature related Hall effects is an open question, which, if realized, would bring a paradigm shift in physical understanding and material selection of Hall effects.

In this work, based on symmetry analysis and a low-energy k·p model, we present a mechanism which is capable of guaranteeing layertronics with quaternary physics, proving the concept of quadri-layertronics. The quadri-layertronics is related to the coupling of out-of-plane ferroelectricity and valley physics in antiferromagnetic multiferroic quadrilayer. The coupling generates the layer-locked Berry curvature and layer-polarized Hall effect, which forces the carriers with four different layer physics to deflect in specific directions. Intriguingly, by controlling the interlayer dipole arrangements via a gate voltage, the quadri-layertronics can be generated and manipulated, enabling the selective induction and detection of layer Hall effect in specific layers. Furthermore, we also validate the gate control of quadri-layertronics in multiferroic antiferromagnet of quadrilayer $OsCl_2$ on the basis of first principles calculations. Our work sheds light on the gate control of quadri-layertronics, which broadens the scope of layertronics application.

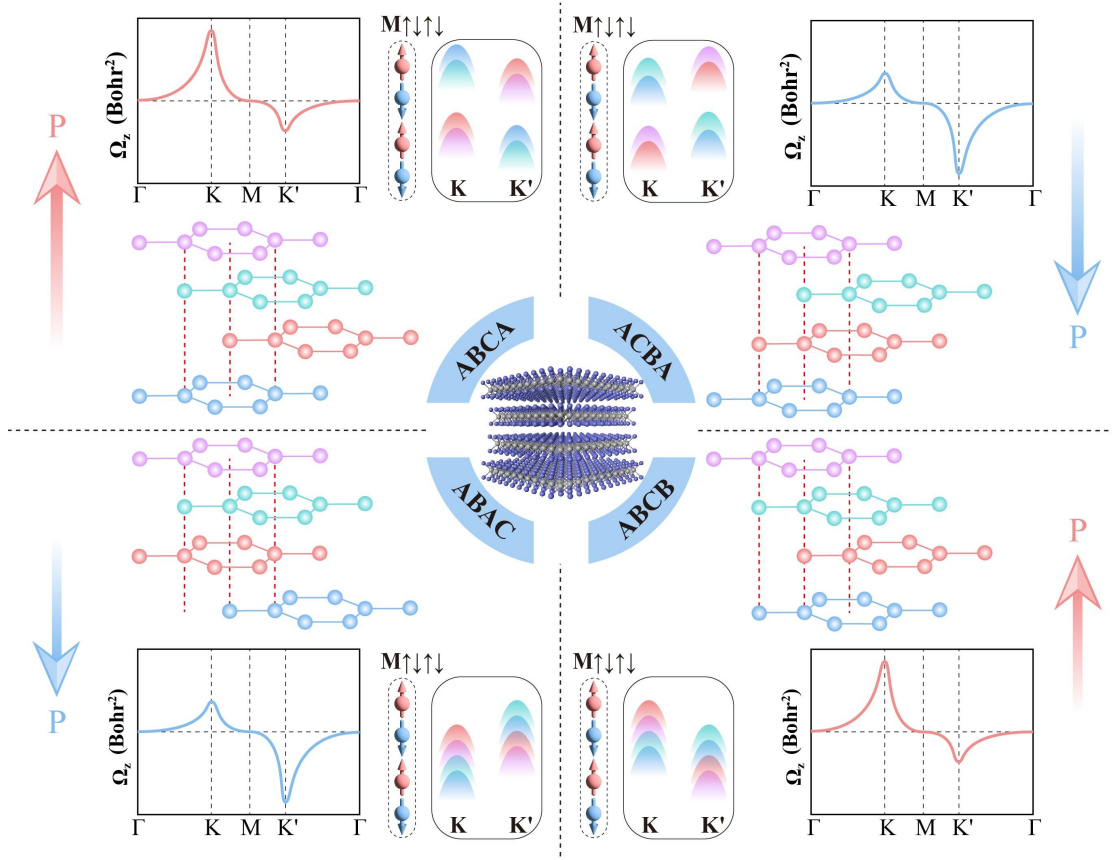

**Fig. 1** Schematic diagrams of ABCA, ACBA, ABAC and ABCB stacked configurations of quadrilayer lattice, with the corresponding Berry curvature, low-energy band spectra, and spin direction. Purple, green, red, blue cones correspond to bands from the α, β, γ and δ layer, respectively. The large arrows labeled on the sides represent the polarization directions.

**Results and Discussion**

The mechanism starts from AAAA stacked hexagonal quadrilayer lattice, wherein the each layer features FM exchange interaction with *P* symmetry broken and spontaneous valley polarization. In such van der Waals lattices, the interlayer exchange interaction is usually dominated by AFM coupling [29,30]. As a result, the spins in the α, β, γ, δ layers are aligned in the sequence of ↑↓↑↓. This fact, along the protection of $M_z$ symmetry, would suppress the overall valley polarization of the AAAA quadrilayer lattice; see **Fig. S1**. Specifically, the K' valley originates from the nearly degenerate spin-up bands of α and γ layers, while the K valley is from the nearly degenerate spin-down bands of β and δ layers. It should be noted that due to the differences in interlayer environment, these bands are not perfectly degenerate, but experience slight splitting. The AAAA quadrilayer lattice shows significant Berry curvatures with opposite signs at the K and K' valleys. It suggests the layer-locked spin and Berry curvatures, i.e., effectively confining these physics to specific layers.

Under interlayer sliding, the AAAA quadrilayer lattice can be transformed into ABCA, ACBA, ABAC, and ABCB stacked configurations. Due to the absence of $M_z$ symmetry, out-of-plane electric polarization would emergence in these four condigurations. ABCA configuration features sequential AB stacking across each two neighboring layers, naturally leading to AA stacking between the α and δ layers and an electric polarization pointing along +z direction. Concerning ACBA configuration, it can be obtained from imposing $M_z$ mirror operation on ABCA configuration, resulting in an opposite out-of-plane electric polarization. Obviously, ABCA and ACBA configurations correspond to two energetically degenerate ferroelectric states with opposite polarizations. Similarly, ABCB configuration involves AB stacking for α-β and β-γ neighboring layers, with AA stacking between the β and δ layers. This suggests an out-of-plane electric polarization along +z direction. ABAC configuration can be derived from ABCB under $M_z$ mirror operation, also resulting in the reversal of out-of-plane polarization. Consequently, ABAC and ABCB form another pair of energetically degenerate ferroelectric states with opposite polarizations. It is interesting to note that arising from different stacking arrangements, the absolute value of electric polarization would be significantly different from that of ABCA and ABAC, indicating they can be transferred into each other through gate-voltage. In addition, considering the absence of $M_z$ symmetry, the valley degeneracy preserved in AAAA configuration would be deformed in ABCA, ACBA, ABAC, and ABCB stacked configurations. Importantly, from the corresponding illustrations of low-energy spectra shown in **Fig. 1**, we can expect the selective manipulation of Berry curvature at specific layers, allowing for gate control of layer-locked physics in all the four constituent layers, i.e., the gate control of quadri-layertronics.

To confirm the proposed mechanism, we construct an effective *k·p* model to describe the low-energy band dispersions near the Fermi level for ABCA, ACBA, ABAC and ABCB configurations. For simplicity, we ignore the interlayer interaction beyond neighboring layers. The Hamiltonian of the model is expressed as:

$$H_k = I_2 \otimes \begin{pmatrix} H_1 & H_{1-2} & 0 & 0 \\ H_{1-2}^\dagger & H_2 & H_{2-3} & 0 \\ 0 & H_{2-3}^\dagger & H_3 & H_{3-4} \\ 0 & 0 & H_{3-4}^\dagger & H_4 \end{pmatrix}.$$

Here, $H_i$ represents the Hamiltonian of the *i*-th layer. And $H_{i-j}$ represents the coupling interaction between the *i*-th and *j*-th layers, which is defined as:

$$H_{i-j} = \begin{pmatrix} t_{cc\_i-j} & 0 \\ 0 & t_{vv\_i-j} \end{pmatrix}.$$

Here, $t_{cc\_i\text{-}j}$ and $t_{vv\_i\text{-}j}$ represent the interlayer hopping energies within the conduction and valence bands, respectively. $H_i$ is the Hamiltonians of the $i$-th layer of the lattice, which can be written as:

$$H_i = H_0 + H_{SOC} + \sigma_i H_{ex} + H_E,$$

Here,

$$H_i^0 = \begin{pmatrix} \frac{\Delta}{2} + \varepsilon + t_{11}'(q_x^2 + q_y^2) & t_{12}(\tau q_x - iq_y) + t_{12}'(\tau q_x + iq_y)^2 \\ t_{12}(\tau q_x + iq_y) + t_{12}'(\tau q_x - iq_y)^2 & -\frac{\Delta}{2} + \varepsilon + t_{22}'(q_x^2 + q_y^2) \end{pmatrix},$$

$$\sigma_i = (-1)^{i+1}.$$

$\triangle$ denotes the band gap at the K (K') valley, $\varepsilon$ refers to the correction energy, $\tau = \pm 1$ corresponds to the valley index, $\vec{q} = \vec{k} - \vec{K}$ represents the momentum vector of electrons relative to the K (K') valley, $t_{12}$ is the effective intralayer nearest-neighbor hopping integral, and $t_{11}'$, $t_{12}'$, and $t_{22}'$ are the second-nearest-neighbor intralayer hopping. The second term $H_{SOC}$ arising from the spin-orbit coupling (SOC) effect can be described as $H_{SOC} = \begin{pmatrix} \tau s \lambda_c & 0 \\ 0 & \tau s \lambda_v \end{pmatrix}$, where the spin index $s = \pm 1$ indicates the spin orientation (up or down), and $\lambda_{c(v)} = E_{c(v)\uparrow} - E_{c(v)\downarrow}$ refers to the spin splitting at the conduction band edge (or valence band edge) in a single layer due to SOC coupling. The third term $H_{ex}$ represents the intrinsic exchange interaction associated with magnetic ions, which can be formulated as $H_{ex} = \begin{pmatrix} -sM_c & 0 \\ 0 & -sM_v \end{pmatrix}$. $M_{c(v)} = E_{c(v)\downarrow} - E_{c(v)\uparrow}$ denotes the effective exchange splitting at the band edge.

The fourth term $H_E$ represents the contribution from interlayer polarization, which is influenced by the specific stacking configuration. Specifically, in ABCA configuration, all interlayer polarizations align in the +z direction, i.e., ↑↑↑. Conversely, in ACBA configuration, all interlayer polarizations align in the -z direction (↓↓↓). ABCB and ABAC configuration shows mixed polarization patterns of ↑↑↓ and ↑↓↓, respectively. These distinct polarizations give rise to different $H_E$ matrices for each configuration, as detailed below:

$$H_{E\_ABCA} = I_2 \otimes \begin{pmatrix} 3U & 0 & 0 & 0 \\ 0 & 2U & 0 & 0 \\ 0 & 0 & U & 0 \\ 0 & 0 & 0 & 0 \end{pmatrix},$$

$$H_{E\_ACBA} = I_2 \otimes \begin{pmatrix} 0 & 0 & 0 & 0 \\ 0 & -U & 0 & 0 \\ 0 & 0 & -2U & 0 \\ 0 & 0 & 0 & -3U \end{pmatrix},$$

$$H_{E\_ABCB} = I_2 \otimes \begin{pmatrix} 2U & 0 & 0 & 0 \\ 0 & U & 0 & 0 \\ 0 & 0 & 0 & 0 \\ 0 & 0 & 0 & -U \end{pmatrix},$$

$$H_{E\_ABAC} = I_2 \otimes \begin{pmatrix} U & 0 & 0 & 0 \\ 0 & 0 & 0 & 0 \\ 0 & 0 & -U & 0 \\ 0 & 0 & 0 & -2U \end{pmatrix}.$$

The *k·p* model results for the low-energy band dispersions near the K and K' valleys in the ABCA, ACBA, ABAC, and ABCB configurations are illustrated in **Fig. S2**. Evidently, these four configurations all exhibit spontaneous valley polarization. For ABCA (ACBA) configuration, the valence band maximum (VBM) at the K (K') valley arises from the δ (α) layer, while for the ABAC (ABCB) configuration, the VBM at the K' (K) valley is contributed by the β (γ) layer. These results agree with the above symmetry analysis, validating the feasibility of the proposed mechanism. Based on these results, the layer physics in layertronics can be engineered into quaternary mode, confirming the concept of quadri-layertronics. For instance, with engineering into ABCA configuration, when the Fermi level is adjusted between the K and K' valleys in the valence band, the nonzero Berry curvature in δ layer acts as an effective magnetic field in the presence of an in-plane electric field. As a result, Bloch electrons acquire an anomalous transverse velocity ($v \sim E \times \Omega(k)$), causing the spin-up holes in the δ layer to accumulate at its left boundary in presence of an in-plane electric field, realizing the LP-AHE. When shifting ABCA configuration into ACBA configuration by interlayer sliding through electric field, the spin-down holes in the α layer will move with opposite transverse velocity due to the reversed Berry curvatures at the K' valleys, accumulating at its right side. Upon transforming into ABAC and ABCB configurations, a similar process occurs, with the spin-down holes accumulating on the right side of the β layer in the ABAC configuration and the spin-up holes accumulating on the opposite side of the γ layer in the ABCB configuration. Thus, the proposed quadri-layertronics can be manipulated by controlling the interlayer dipole arrangements via a gate voltage.

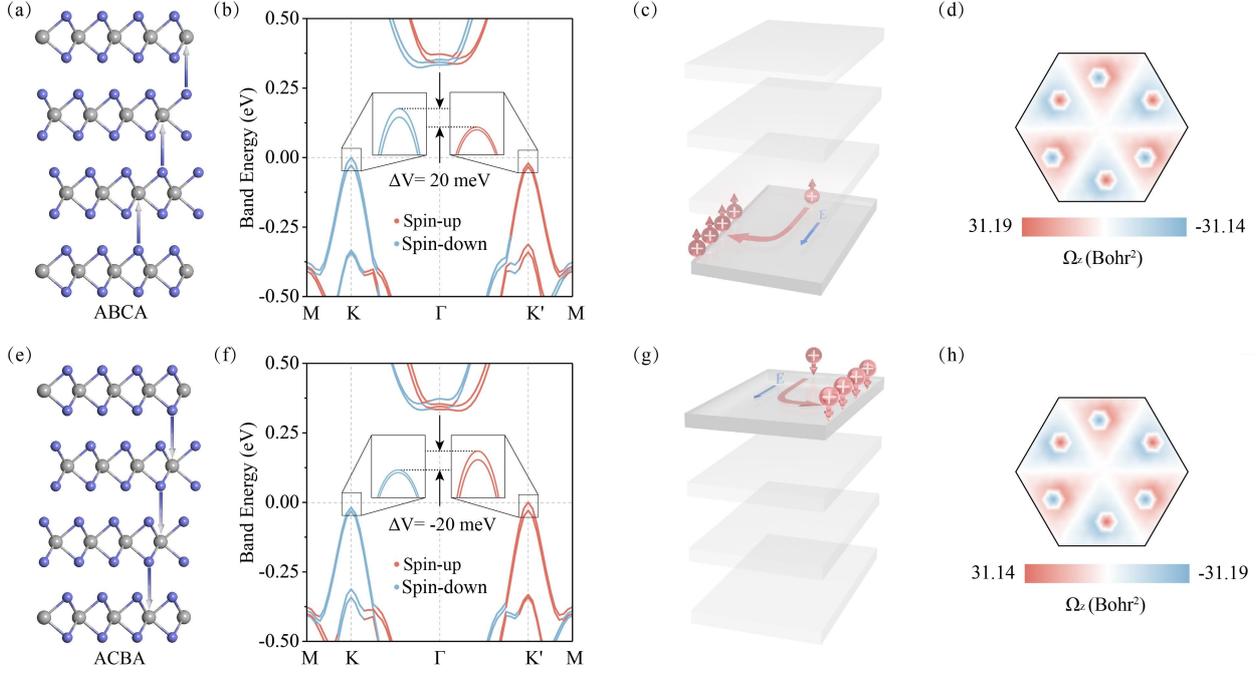

**Fig. 2** Crystal structures of (a) ABCA and (c) ACBA configurations of quadrilayer OsCl$_2$. Band structures and enlarged low-energy band dispersions, diagrams of the LP-AHE under hole doping and Berry curvatures of (b, c and d) ABCA and (f, g and h) ACBA configurations. The Fermi level is set to 0 eV.

Next, we discuss the realization of quadri-layertronics in real materials of quadrilayer OsCl$_2$. **Fig. S3(a-c)** illustrate the crystal structure of single-layer OsCl$_2$, which exhibits a hexagonal lattice formed by one Os atomic layer sandwiched between two Cl atomic layers. It shows the space group $P\bar{6}m2$, indicating the P-symmetry breaking. The thermal and dynamical stability of single-layer OsCl$_2$ are confirmed with phonon calculations and molecular dynamics simulations, as shown in **Fig. S4**. The trigonal prismatic crystal field acting on the Os atom causes its five d orbitals to split into $a$ ($d_{z2}$), $e_1$ ($d_{x2-y2}$, $d_{xy}$), and $e_2$ ($d_{xz}$, $d_{yz}$) orbitals, with $a$ having the lowest energy, followed by the $e_1$ and $e_2$ orbitals, as depicted in **Fig. S3(e)**. After the Os atom donates two electrons to the six coordinated Cl atoms, the remaining six electrons half-fill the $e_1$ and $e_2$ orbitals, resulting in a magnetic moment of 4 $\mu_B$. This is consistent with previous work[31,32]. After considering different magnetic configurations, the FM coupling is found to be the ground state. The calculated band structures of single-layer OsCl$_2$ with SOC are shown in **Fig. S4(b)**. Its VBM is located at the K' valley, and the spontaneous valley polarization in the valence band reaches 215.21 meV.

Based on the proposed mechanism, we construct ABCA, ACBA, ABAC and ABCB configurations of OsCl$_2$, as illustrated in **Fig. 2(a, e)** and **3(a, e)**. To explore the ground state exchange interactions in quadrilayer OsCl$_2$, we consider different magnetic configurations. Our calculations reveal that the intralayer and interlayer exchange interactions of all these four configurations are dominated by FM and AFM couplings, respectively, resulting in A-type AFM order; see **Table S1**. This indicates that the spin in α, β, γ, δ layers are aligned in the sequence of M↑↓↑↓, satisfying the proposed conditions. Additionally, the magnetocrystalline anisotropy energy calculations reveal the preference for out-of-plane magnetization for all four configurations; see **Table S1**. Due to the lack of P and $M_z$ symmetries, spontaneous out-of-plane electric polarizations are induced. For ABCA and ACBA configurations, the plane averaged electrostatic potential reveals a distinct polarization of $\Delta V$ = -0.033 eV in ABCA (out-of-plane electric polarization along +z) and $\Delta V$ = 0.033 eV in ACBA (out-of-plane electric polarization along -z); see **Fig. 4(b)**. Different from ABCA and ACBA configurations, ABAC and ABCB configurations exhibit different polarization patterns, with ABAC pattern showing $\Delta V$ = 0.015 eV and ABCB pattern showing $\Delta V$ = -0.015 eV; see **Fig. 4(d)**. The calculated electric polarizations using the Berry phase method are 0.25 and -0.25 pC/m for ABCA and ACBA, and -0.12 pC/m and 0.12 pC/m for ABAC and ABCB, respectively. Therefore, these four ferroelectric states of quadrilayer $OsCl_2$ can be generated and manipulated by controlling gate voltage.

**Fig. 2(b, f)** and **Fig. 3(b, f)** present the band structures of ABCA, ACBA, ABAC and ABCB configurations of quadrilayer $OsCl_2$. Both ABCA and ACBA exhibit an indirect band gap of 0.35 eV, with the VBM lying at the K and K' points, respectively. ABAC and ABCB configurations show an indirect band gap of 0.61 eV, with the VBM locating at the K' and K points, respectively. The out-of-plane electric polarization in these structures plays a crucial role in determining the spin and layer physics of the VBM. Specifically, in ABCA (ACBA) pattern, the VBM arises from the spin-down (spin-up) channel of the δ (α) layer, leading to spontaneous valley polarization that is layer-locked. Similarly, ABAC (ABCB) configurations exhibit analogous behavior, where the VBM in ABAC (ABCB) stems from the spin-up (spin-down) channel of the β (γ) layer. The calculated spontaneous valley polarizations in the valence bands for these four configurations are ±20 meV and ±27 meV, respectively. These values are significantly larger than the $WSe_2/CrI_3$ (3.5 meV)[33] and $WSe_2/EuS$ (2.5 meV)[34] heterojunctions in experiments.

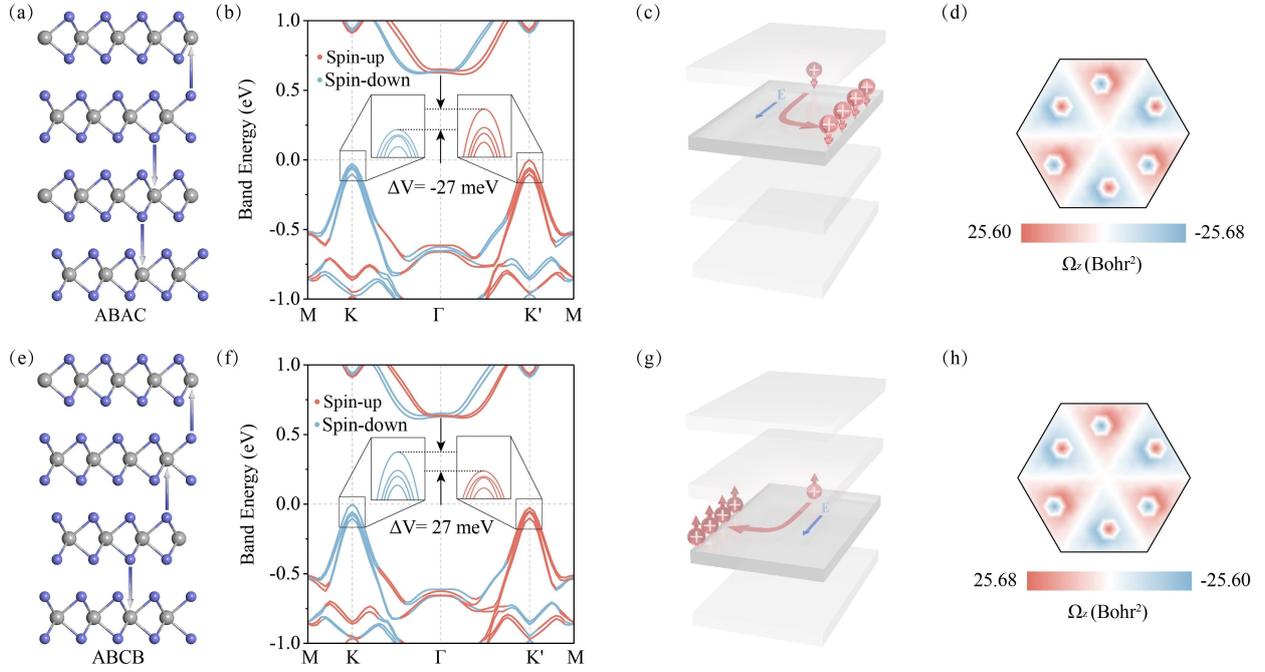

**Fig. 3** Crystal structures of (a) ABAC and (c) ABCB configurations of quadrilayer OsCl$_2$. Band structures and enlarged low-energy band dispersions, diagrams of the LP-AHE under hole doping and Berry curvatures of (b, c and d) ABAC and (f, g and h) ABCB configurations. The Fermi level is set to 0 eV.

After identifying the spontaneous valley polarization in quadrilayer OsCl$_2$, we investigate the Berry curvatures associated with layer degree. We employed the VASPBERRY code to calculate their Berry curvatures in ABCA, ACBA, ABAC and ABCB stacked quadrilayer OsCl$_2$, which is defined as[35]:

$$\Omega(k) = -\sum_n \sum_{n \neq n'} f_n \frac{2Im\langle\psi_{nk}|v_x|\psi_{n'k}\rangle\langle\psi_{n'k}|v_y|\psi_{nk}\rangle}{(E_n - E_{n'})^2}.$$

Here, $f_n$ is the Fermi-Dirac distribution function, $\psi_{nk}$ is the Bloch wave function with eigenvalue $E_n$, and $v_x$ ($v_y$) is the velocity operator along $x(y)$ direction. **Fig. 2(d, h)** and **3(d, h)** plot the calculated Berry curvatures for the four configurations. The Berry curvature at the K (K') valley in the ABCA (ACBA) configuration shows a large positive (negative) value. Since the valleys are locked to layers, the layer-locked Berry curvature is realized. Since the Berry curvature is layer-locked, the carriers in a given layer would acquire an anomalous velocity under an in-plane electric field and transversely move to one side of this layer. In the ABCA configuration, spin-up holes in the K valley acquire an anomalous velocity $v_a = -\frac{e}{\hbar}E \times \Omega(k)$[35] under an in-plane electric field $E$, and transversely move to the left edge of the δ layer, as shown in **Fig. 2(c)**. Conversely, in ACBA configuration, spin-down holes in the K' valley gain a reversed anomalous velocity of $-v_a$, and accumulate at the right edge of the α layer, as shown in **Fig. 2(g)**. ABAC and ABCB configurations exhibit analogous behaviors with respect to their Berry curvatures. In ABAC configuration, the negative Berry curvature at the K' valley generates an anomalous velocity -$v_a$, which directs spin-down holes toward the right edge of the β layer; see **Fig. 3(c, d)**. While in ABCB configuration, the positive Berry curvature at the K valley results in a reversed velocity $v_a$, pushing spin-up holes to the left edge of the γ layer; see **Fig. 3(g, h)**. Consequently, the layertronics with quaternary physics is observed in quadrilayer OsCl$_2$, with each configuration enabling the LP-AHE in a different layer, resulting in the emergence of quadri-layertronics.

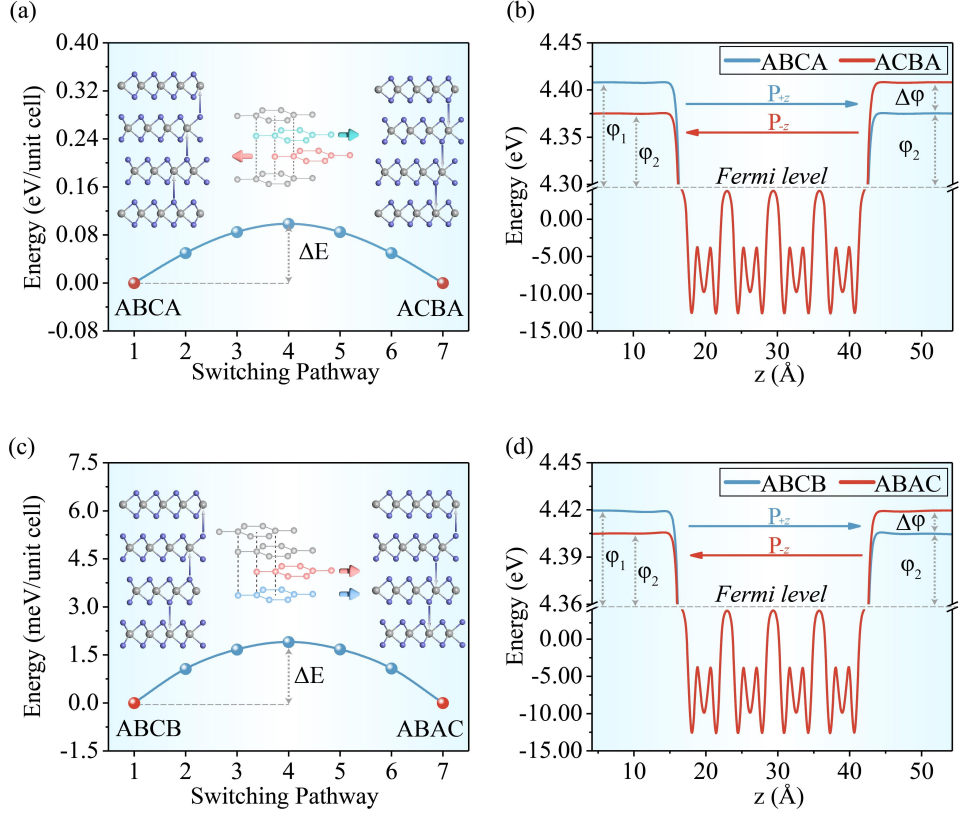

**Fig. 4** Energy profiles for ferroelectric switching between (a) ABCA→ACBA and (c) ABCB→ABAC configurations in quadrilayer OsCl$_2$. Insets show the crystal structure transitions. (b) Plane-averaged electrostatic potentials for ABCA and ACBA configurations along the *z* direction. (d) Plane-averaged electrostatic potentials for ABCB and ABAC configurations along the *z* direction.

At last, we discuss the possibility for gate control of quadri-layertronics in quadrilayer OsCl$_2$. To this end, we investigate the switching between the quaternary states. As illustrated in **Fig. 4(a)**, the switching between ABCA and ACBA configurations is realized through sliding the β layer of the OsCl$_2$ along the $\left[-\frac{2}{3}, -\frac{1}{3}, 0\right]$, $\left[\frac{1}{3}, -\frac{1}{3}, 0\right]$ or $\left[\frac{1}{3}, \frac{2}{3}, 0\right]$ direction, while sliding the γ layer in the opposite direction of $\left[\frac{2}{3}, \frac{1}{3}, 0\right]$, $\left[-\frac{1}{3}, \frac{1}{3}, 0\right]$ or $\left[-\frac{1}{3}, -\frac{2}{3}, 0\right]$. Along with this switching, the polarization direction is also reversed. The energy barrier associated with this switching is 0.098 eV per unit cell; see **Fig. 4(a)**. **Fig. 4(c)** shows the switching between the ABAC and ABCB configurations, which is driven by the concurrent sliding of the γ and δ layers. This interlayer sliding also reverses the polarization, with an associated energy barrier of 1.9 meV per unit cell. The relatively low energy barriers for these transformations can be readily overcome by external stimuli of electric fields. This indicates that the quadri-layertronics in quadrilayer $OsCl_2$ can be generated and manipulated by controlling the interlayer dipole arrangements via a gate voltage, enabling the selective induction and detection of layer Hall effect in specific layers. Finally, it is worth noting that the proposed mechanism could be extended to other van der Waals multilayers.

**Conclusions**

In conclusion, we reveal the layer physics in layertronics can be engineered into quaternary mode, and propose the concept of quadri-layertronics. We show that the quadri-layertronics can be generated and manipulated by tuning the stacking arrangements via a gate voltage, guaranteeing the selective induction and detection of layer Hall effect in specific layers. Furthermore, we also validate the gate control of quadri-layertronics in multiferroic antiferromagnet of quadrilayer $OsCl_2$ using first principles calculations.

**AUTHOR INFORMATION**

Corresponding Authors

*E-mail: sps_zhangt@ujn.edu.cn (T.Z.).

*E-mail: yandong.ma@sdu.edu.cn (Y.M.).

ORCID

Ting Zhang: 0009-0008-5670-9955

Yandong Ma: 0000-0003-1572-7766

Notes

The authors declare no competing financial interest.

**SUPPORTING INFORMATION**

Supporting Information Available: details regarding the method, the *k·p* model, and the properties of single-layer $OsCl_2$.

ACKNOWLEDGMENTS

T. Z. and M. W. contributed equally to this work. This work was supported by the Natural Science Foundation of Shandong Province (No. ZR2024QA051) National Natural Science Foundation of China (Nos. 12274261 and 12074217) and Taishan Young Scholar Program of Shandong Province.